\documentclass[aip,jap,reprint]{revtex4-1}%
\usepackage{graphicx}
\usepackage{dcolumn}
\usepackage{bm}
\usepackage{amsmath}
\usepackage{pslatex}%
\usepackage{amsfonts}%
\usepackage{amssymb}

\begin{document}
\title
{Modeling bound-to-continuum terahertz quantum cascade lasers: The role of Coulomb interactions}
\author{Christian Jirauschek}
\email{jirauschek@tum.de}
\homepage{http://www.nano.ei.tum.de/noether}
\affiliation
{Emmy Noether Research Group ``Modeling of Quantum Cascade Devices``,Technische
Universit\"{a}t M\"{u}nchen, D-80333 Munich, Germany}
\affiliation{Institute for Nanoelectronics, Technische Universit\"{a}%
t M\"{u}nchen, D-80333
Munich, Germany}
\author{Alpar Matyas}
\affiliation
{Emmy Noether Research Group ``Modeling of Quantum Cascade Devices``,Technische
Universit\"{a}t M\"{u}nchen, D-80333 Munich, Germany}
\affiliation{Institute for Nanoelectronics, Technische Universit\"{a}%
t M\"{u}nchen, D-80333
Munich, Germany}
\author{Paolo Lugli}
\affiliation{Institute for Nanoelectronics, Technische Universit\"{a}%
t M\"{u}nchen, D-80333
Munich, Germany}
\date{\today, published as J. Appl. Phys. 107, 013104 (2010)}
\begin{abstract}
Based on an ensemble Monte Carlo analysis, we show that Coulomb interactions play a dominant role in bound-to-continuum terahertz quantum cascade lasers and thus require careful modeling. Coulomb interactions enter our simulation in
the form of space charge effects as well as Coulomb scattering events. By comparison to a full many-subband Coulomb
screening model, we show that simplified approaches produce considerable deviations for such structures. Also the spin
dependence of electron-electron scattering has to be adequately considered. Moreover, we demonstrate that iterative
Schr{\"o}%
dinger-Poisson and carrier transport simulations are necessary to correctly account for space charge effects.
\end{abstract}
\maketitle

\section{\label{sec:level1}INTRODUCTION}

For terahertz quantum cascade lasers (QCLs), two types of structures play a
major role: the resonant-phonon (RP) and the bound-to-continuum (BTC) design.
In RP structures, efficient depletion of the lower laser level is achieved by
tuning the corresponding transition to the longitudinal-optical (LO) phonon
energy (36 meV in GaAs). By contrast, BTC designs are based on minibands,
enhancing the influence of Coulomb interactions in two ways. First, the close
energetic spacing of the miniband levels favors electron-electron (e-e) over
LO phonon scattering in the carrier transport. Also, the large spatial extent
of the minibands across many wells, together with the localization of the
positively charged donors in typically a single well, leads to considerable
conduction band bending due to space charge effects.\cite{2005ApPhL..86u1117J}
Thus, a careful modeling of Coulomb interactions is necessary to analyze
carrier transport in BTC QCLs, and here specifically the role of Coulomb
interactions in such structures. However, e-e scattering is much more
computationally demanding than single-electron processes like electron-phonon
interactions, hampering its inclusion in quantum mechanical simulations of
QCLs beyond the mean-field approximation.\cite{2009PhRvB..79s5323K} The
numerical load is further increased by the large spatial extent of the
minibands. Due to its efficiency, the semiclassical ensemble Monte-Carlo (EMC)
method\ is well-suited for investigating BTC structures. However, EMC
simulations have up to now typically focused on terahertz RP
structures,\cite{2001ApPhL..78.2902I,2003ApPhL..83..207C,2004ApPhL..84..645C,2005JAP....97d3702B,2006ApPhL..89u1115L,2007JAP...101h6109J,2008pssc..5..221J}
while only few results are available for equivalent BTC or related chirped
superlattice designs.\cite{2001ApPhL..79.3920K,2009JAP...105l3102J}

We present an EMC simulation tool optimized for the simulation of terahertz
BTC structures. We employ a full many-subband (MS) screening model for
electron-electron scattering and include the exchange effect for the
parallel-spin collisions. Space charge effects are adequately considered by
performing iterative Schr\"{o}dinger-Poisson (SP) and carrier transport
simulations, yielding so-called self-self-consistent
solutions.\cite{2005ApPhL..86u1117J} This tool allows us to properly assess
the role of Coulomb interactions for the carrier transport and gain in a
typical terahertz BTC design. Widely-used approximations, such as
non-iterative simulations, using simplified screening models, or neglecting
the spin dependence of e-e scattering, usually work well for RP structures,
but can lead to significant deviations for terahertz BTC QCLs, as shown in
this paper.

\section{METHOD}

The simulation tool, consisting of a three-dimensional EMC and an SP solver,
allows for a self-consistent analysis of the carrier transport and optical
gain in the QCL structure.\cite{2009JAP...105l3102J} The SP solver yields the
subband eigenenergies and wave functions, needed as an input for the
semiclassical EMC carrier transport simulation. The obtained electron
distribution in the structure gives rise to space charge effects, resulting in
conduction band bending and altered subbands. Thus, iterative runs of SP\ and
EMC simulations are necessary to obtain convergence, corresponding to
self-self-consistent solutions.\cite{2005ApPhL..86u1117J}

All essential scattering mechanisms are accounted for in the carrier transport
simulation. Included is elastic electron-impurity (e-i) and interface
roughness scattering, as well as inelastic interactions of electrons with
acoustic and longitudinal-optical (LO) phonons. Nonequilibrium phonon effects
are also taken into account.\cite{2006ApPhL..88f1119L} Being evaluated as a
two-electron process, e-e scattering assumes a special role in the EMC
simulation and has the highest computational complexity. Its implementation is
more closely discussed in Section \ref{e_e}. For all scattering processes,
Pauli's exclusion principle is taken into account.\cite{1985Lugli_excl}
Periodic boundary conditions are employed, i.e., electrons leaving the device
on one side are automatically injected into the equivalent subband on the
opposite side.\cite{2001ApPhL..78.2902I}

Coulomb interactions enter our simulation in the form of space charge effects
as well as individual e-e and e-i scattering events. As discussed above, space
charge effects and e-e scattering play an especially important role in THz BTC
structures due to the formation of minibands. Thus, the implementation of
these mechanisms will be discussed in detail in the following.

\subsection{Space charge effects}

The subband wave functions $\psi_{n}\left(  z\right)  $ and eigenenergies
$E_{n}$ for the simulated QCL structure are obtained by solving the SP
system\cite{2009IJQE...45..1059J}
\begin{subequations}
\label{sp}%
\begin{align}
\left(  -\frac{\hbar^{2}}{2}\partial_{z}\frac{1}{m^{\ast}}\partial_{z}%
+V-E_{n}\right)  \psi_{n}  &  =0,\label{sp1}\\
-\varepsilon\partial_{z}^{2}\varphi-e\left(  N-\sum_{n}n_{\mathrm{2D}%
,n}\left|  \psi_{n}\right|  ^{2}\right)   &  =0. \label{sp2}%
\end{align}
\end{subequations}
Here, a position independent permittivity $\varepsilon$ is assumed.
Furthermore, $m^{\ast}\left(  z\right)  $\ and $N\left(  z\right)  $ are the
electron effective mass and doping concentration of the structure, $e$ and
$\hbar$ are the elementary charge and the reduced Planck constant, and
$n_{\mathrm{2D},n}$ is the electron sheet density of level $n$.\ The
self-consistent potential is given by $V\left(  z\right)  =V_{0}%
(z)-e\varphi\left(  z\right)  $, where $V_{0}$ is the conduction band profile
and $\varphi$ is the electric potential due to the space charge profile. For
QCLs, the subband energies and wave functions are typically computed for a
single central period of the structure; the subbands in other periods are then
obtained by appropriate shifts of the solutions in energy and position. For
the first run of the SP solver, a thermal occupation of the subbands according
to Fermi-Dirac statistics is assumed.\cite{2009IJQE...45..1059J} To obtain
self-self-consistent solutions, the subband occupations are subsequently
extracted from the EMC analysis, which is carried out alternately with the
SP\ simulation until mutual convergence is obtained.

\subsection{\label{e_e}Electron-electron scattering}

In the EMC simulation, e-e scattering is implemented as a two-electron
process.\cite{1988PhRvB..37.2578G,1995PhRvB..5116860M} An electron in an
initial state $\left|  i\mathbf{k}\right\rangle $, i.e., subband $i$ and
in-plane wave vector $\mathbf{k}$, scatters to a final state $\left|
j\mathbf{k}^{\prime}\right\rangle $, accompanied by a transition of a second
electron from a state $\left|  i_{0}\mathbf{k}_{0}\right\rangle $ to $\left|
j_{0}\mathbf{k}_{0}^{\prime}\right\rangle $. The total scattering rate from
$\left|  i\mathbf{k}\right\rangle $ to a subband $j$ is then obtained by the
Fermi golden rule,
\begin{equation}
R_{i\mathbf{k}\rightarrow j}=\frac{m^{\ast}}{4\pi\hbar^{3}A}\sum_{i_{0}%
,j_{0},\mathbf{k}_{0}}f_{i_{0}}\left(  \mathbf{k}_{0}\right)  \int_{0}^{2\pi
}d\theta\left|  M_{ii_{0}jj_{0}}\left(  Q\right)  \right|  ^{2}, \label{R_ee}%
\end{equation}
with the electron effective mass $m^{\ast}$, cross section area $A$ and
carrier distribution function $f_{i_{0}}$ in subband $i_{0}$. $\theta$ is the
angle between $\mathbf{g}=\mathbf{k}_{0}-\mathbf{k}$ and $\mathbf{g}^{\prime
}=\mathbf{k}_{0}^{\prime}-\mathbf{k}^{\prime}$, and $\mathbf{Q=k-k}^{\prime}$
(with $Q=\left|  \mathbf{Q}\right|  $) denotes the exchanged wavevector.

Different approaches with varying degrees of complexity exist to compute the
transition matrix element $M_{ii_{0}jj_{0}}$ from the\ bare Coulomb matrix
elements,%
\begin{align}
V_{ii_{0}jj_{0}}^{\mathrm{b}}\left(  Q\right)   &  =\frac{e^{2}}{2\varepsilon
Q}\int_{-\infty}^{\infty}dz\int_{-\infty}^{\infty}dz^{\prime}\left[  \psi
_{i}\left(  z\right)  \psi_{i_{0}}\left(  z^{\prime}\right)  \right.
\nonumber\\
&  \times\left.  \psi_{j}\left(  z\right)  \psi_{j_{0}}\left(  z^{\prime
}\right)  \exp\left(  -Q\left|  z-z^{\prime}\right|  \right)  \right]  .
\label{Vbare}%
\end{align}
First, the screened Coulomb matrix elements $V_{ii_{0}jj_{0}}^{\mathrm{s}%
}\left(  Q\right)  $ are obtained from $V_{ii_{0}jj_{0}}^{\mathrm{b}}\left(
Q\right)  $ by applying a more or less sophisticated screening model. In the
random phase approximation (RPA), they are found by solving the equation
system\cite{1999PhRvB..5915796L}%
\begin{equation}
V_{ii_{0}jj_{0}}^{\mathrm{s}}=V_{ii_{0}jj_{0}}^{\mathrm{b}}+\sum_{mn}%
V_{imjn}^{\mathrm{b}}\Pi_{mn}V_{mi_{0}nj_{0}}^{\mathrm{s}}. \label{Vscr}%
\end{equation}
Here, $\Pi_{mn}\left(  Q\right)  $ is the polarizability tensor, given in the
long wavelength limit ($Q\rightarrow0$) by%

\begin{equation}
\Pi_{mn}=\left\{
\begin{array}
[c]{cc}%
\frac{n_{\mathrm{2D},m}-n_{\mathrm{2D},n}}{E_{m}-E_{n}}, & m\neq n,\\
-\frac{m^{\ast}}{\pi\hbar^{2}}f_{n}\left(  0\right)  , & m=n.
\end{array}
\right.
\end{equation}
For collisions of electrons with parallel spin,\ interference\ occurs between
$V_{ii_{0}jj_{0}}^{\mathrm{s}}\ $and the 'exchange' matrix element
$V_{ii_{0}j_{0}j}^{\mathrm{s}}$.\cite{1994SeScT...9..478M} Accounting for this
exchange effect, the magnitude squared of the transition matrix element
$M_{ii_{0}jj_{0}}$ is then given
by\cite{1995PhRvB..5116860M,1994SeScT...9..478M}%
\begin{align}
\left|  M_{ii_{0}jj_{0}}\right|  ^{2}  &  =\frac{p_{a}}{2}\left[  \left|
V_{ii_{0}jj_{0}}^{\mathrm{s}}\left(  Q^{-}\right)  \right|  ^{2}+\left|
V_{ii_{0}j_{0}j}^{\mathrm{s}}\left(  Q^{+}\right)  \right|  ^{2}\right]
\nonumber\\
&  +\frac{p_{p}}{2}\left|  V_{ii_{0}jj_{0}}^{\mathrm{s}}\left(  Q^{-}\right)
-V_{ii_{0}j_{0}j}^{\mathrm{s}}\left(  Q^{+}\right)  \right|  ^{2}, \label{M}%
\end{align}
where $p_{a}=p_{p}=1/2$ are the probabilities for antiparallel and parallel
spin collisions, respectively, and%
\begin{equation}
Q^{\pm}=\frac{1}{2}\left[  2g^{2}+g_{0}^{2}\pm2g\left(  g^{2}+g_{0}%
^{2}\right)  ^{1/2}\cos\theta\right]  ^{1/2},
\end{equation}
with $g=\left|  \mathbf{g}\right|  $ and $g_{0}^{2}=4m^{\ast}\left(
E_{i}+E_{i_{0}}-E_{j}-E_{j_{0}}\right)  /\hbar^{2}$.

Commonly, simplified screening models are used to avoid the numerical load
associated with solving Eq. (\ref{Vscr}%
).\cite{2005JAP....97d3702B,2006ApPhL..89u1115L} Furthermore, often the
exchange effect is neglected when calculating $M_{ii_{0}jj_{0}}$%
.\cite{1994SeScT...9..478M} As discussed in Section \ref{sec:level1}, e-e
scattering plays a particularly significant role in BTC structures. Thus, the
validity of such approximations in BTC designs deserves special scrutiny, as
shown below.

\section{RESULTS AND DISCUSSION}

We present simulation results for a $3.5\,\mathrm{THz}$ BTC
design,\cite{2003ApPhL..82.3165S} demonstrating the importance of Coulomb
interactions on the carrier transport and material gain in such structures. In
particular, we investigate the validity of neglecting the exchange effect and
employing simplified screening models for the evaluation of e-e scattering.
Also the importance of self-self-consistent simulations for a proper inclusion
of space charge effects is discussed. In our EMC simulation, all scattering
mechanisms are evaluated self-consistently. Only interface roughness (IR),
whose parameters are hard to measure and depend critically on the growth
conditions, has to be described in terms of a phenomenological
model.\cite{2009JAP...105l3102J,2006PhRvB..73h5311L} For the IR mean height
and correlation length, we use typical values of $\Delta=0.12\,\mathrm{nm}$,
$\Gamma=10\,\mathrm{nm}$.\cite{2009JAP...105l3102J,2008ApPhL..92h1102N}

\subsection{\label{space_charge}Space charge effects}

\begin{figure}[ptb]
\includegraphics{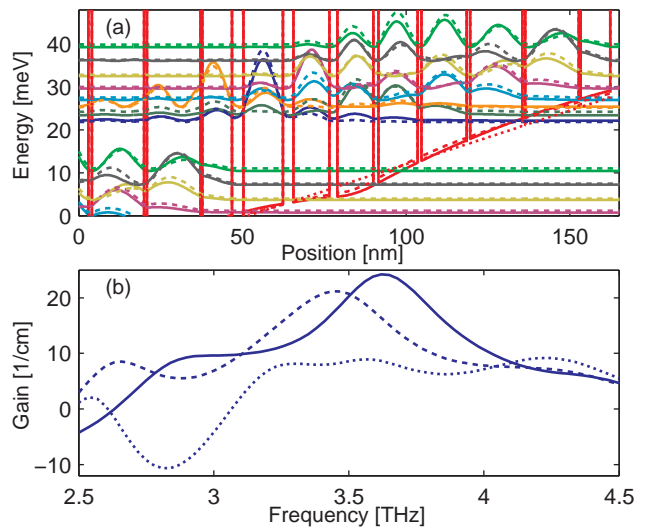}
\caption{(Color online) (a) Conduction band profile and probability densities,
as obtained for thermally occupied subbands (dashed lines) and by a
self-self-consistent simulation (solid lines). For comparison, also the
conduction band profile without space charge effects included is displayed
(dotted line). (b) Simulated spectral gain versus frequency for the cases
shown above.}%
\label{cond}%
\end{figure}

In Fig. \ref{cond}, the simulated conduction band profile and spectral gain of
the investigated BTC QCL is shown for a lattice temperature $T_{\mathrm{L}%
}=10\,\mathrm{K}$ at the design bias of $2.5\,\mathrm{kV/cm}$, where the
maximum gain is obtained in the simulation. Solid lines indicate fully
self-self-consistent results; here, the SP\ and EMC simulations are carried
out iteratively until convergence is obtained. Dashed lines indicate the
results obtained by solving the SP system once in the beginning assuming a
thermal carrier distribution,\cite{2009IJQE...45..1059J} and then performing a
self-consistent EMC simulation, which is a quite common
approach.\cite{2001ApPhL..78.2902I,2005JAP....97d3702B,2007JAP...101h6109J,2008pssc..5..221J}
For comparison, also the conduction band profile and spectral gain is shown as
obtained with deactivated Poisson solver (dotted lines), i.e., when no space
charge effects are included. In all cases, e-e scattering is evaluated taking
into account the exchange effect as well as screening, which is considered in
RPA by repeatedly solving Eq. (\ref{Vscr}) to account for changes in the
carrier distribution during the EMC simulation.

The results shown in Fig. \ref{cond} indicate that the proper inclusion of
space charge effects greatly affects the simulation outcome for the
investigated BTC structure. A significant conduction band bending is observed
for activated Poisson solver (solid and dashed lines in Fig. \ref{cond}(a)),
changing the subband eigenenergies and wave functions. Comparison to
experimental data shows that only the simulation results with space charge
effects included are in line with experiment. The gain profile obtained with
deactivated Poisson solver is exceedingly broad, flat and low, see Fig.
\ref{cond}(b) (dotted curve). The low peak gain of around $9\,\mathrm{cm}%
^{-1}$ at around $4.2\,\mathrm{THz}$ is in sharp contrast to the
experimentally observed lasing of the structure at $3.5\,\mathrm{THz}%
$.\cite{2003ApPhL..82.3165S} Also the large gain bandwidth is not in accord
with electroluminescence measurements, yielding full width at half maximum
(FWHM) widths of clearly below $1\,\mathrm{THz}$.\cite{2003ApPhL..82.3165S}
The inclusion of space charge effects leads to a realistic gain profile
centered around $3.5\,\mathrm{THz}$ in agreement with experiment. The
FWHM\ widths of $0.65\,\mathrm{THz}$ (solid curve) and $0.66\,\mathrm{THz}$
(dashed curve) agree reasonably well with the experimental value of
$0.85\,\mathrm{THz}$ at $10\,\mathrm{K}$, extracted from electroluminescence
measurements.\cite{2003ApPhL..82.3165S} Still, the two gain profiles are
somewhat different, with peak gain values of $24.2\,\mathrm{cm}^{-1}$ for the
self-self-consistent and $21.2\,\mathrm{cm}^{-1}$ for the non-iterative
approach. This illustrates the importance of self-self-consistent simulations
for terahertz BTC designs, where space charge effects tend to play a more
pronounced role than in equivalent RP structures.

\subsection{\label{spin}Exchange effect}

There are two common approaches to implement e-e scattering without explicitly
considering the spin dependence. One method, which tends to overestimate the
exchange effect, is to completely neglect the parallel spin
collisions,\cite{1988PhRvB..37.2578G,1994SeScT...9..478M} implying $p_{a}%
=1/2$, $p_{p}=0$ in Eq. (\ref{M}). Another common approach is to ignore the
spin dependence. Parallel spin collisions are then treated the same way as
antiparallel spin contributions,\cite{1994SeScT...9..478M} which corresponds
to $p_{a}=1$, $p_{p}=0$ in Eq. (\ref{M}). In moderately doped RP\ structures
at elevated temperatures, where the carrier transport is dominated by
LO\ phonon scattering, the contribution of the exchange effect is usually
negligible. In BTC designs, based on minibands with closely spaced energy
levels, e-e scattering plays a more pronounced role.\cite{2009JAP...105l3102J}%
\begin{figure}[ptb]
\includegraphics{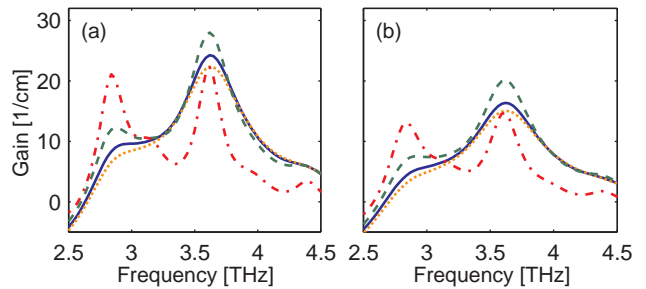}
\caption{(Color online) Simulated spectral gain vs frequency, as obtained by
fully taking into account (solid curves) and ignoring the exchange effect
(dotted curves), and by ignoring parallel spin collisions (dashed curves). For
comparison, the result obtained with no e-e scattering is also shown
(dash-dotted curves). (a) $T_{\mathrm{L}}=10\,\mathrm{K}$; (b) $T_{\mathrm{L}%
}=90\,\mathrm{K}$.}%
\label{g_spin}%
\end{figure}

Figure \ref{g_spin} contains the self-self-consistently simulated gain spectra
at $T_{\mathrm{L}}=10\,\mathrm{K}$ and $T_{\mathrm{L}}=90\,\mathrm{K}$, taking
into account screening in the RPA. Results are shown for $p_{a}=p_{p}=1/2$
(solid curve), $p_{a}=1$, $p_{p}=0$ (dotted curve), $p_{a}=1/2$, $p_{p}=0$
(dashed curve), and $p_{a}=p_{p}=0$ (dash-dotted curve). The last case, which
corresponds to completely neglecting e-e scattering in the simulation, yields
two narrow gain spikes around $2.8$ and $3.6\,\mathrm{THz}$, largely deviating
from the experimental electroluminescence
measurements.\cite{2003ApPhL..82.3165S} This illustrates the importance of e-e
scattering for such structures. As discussed in Section \ref{space_charge},
for $T_{\mathrm{L}}=10\,\mathrm{K}$ the full simulation with exchange effect
included yields a realistic gain profile, see solid curve in Fig.
\ref{g_spin}(a). Ignoring exchange (dotted curve) leads to an overestimation
of the scattering, resulting in a peak gain reduction by $8\%$, and an
increase of the FWHM\ gain width by $11\%$. On the other hand, completely
neglecting parallel spin collisions (dashed curve) leads to a gain width
reduction of $24\%$, and an increase of the peak gain by $15\%$. For
$T_{\mathrm{L}}=90\,\mathrm{K}$, see Fig. \ref{g_spin}(b), the gain gets
somewhat broadened and lowered in agreement with
experiment;\cite{2003ApPhL..82.3165S,2009JAP...105l3102J} also here, ignoring
the exchange effect or parallel spin collisions has similar effects on the
simulated gain profile as discussed for $T_{\mathrm{L}}=10\,\mathrm{K}$.

\begin{figure}[ptb]
\includegraphics{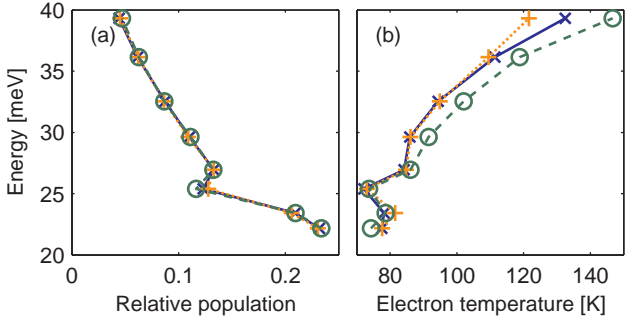}
\caption{(Color online) Simulation results for (a) relative subband
occupations and (b) subband temperatures, as obtained by fully taking into
account (x-marks, solid curves) and ignoring the exchange effect (plus signs,
dotted curves), and by ignoring parallel spin collisions (circles, dashed
curves). The lines are guide to the eye.}%
\label{pT_spin}%
\end{figure}

\begin{figure}[ptb]
\includegraphics{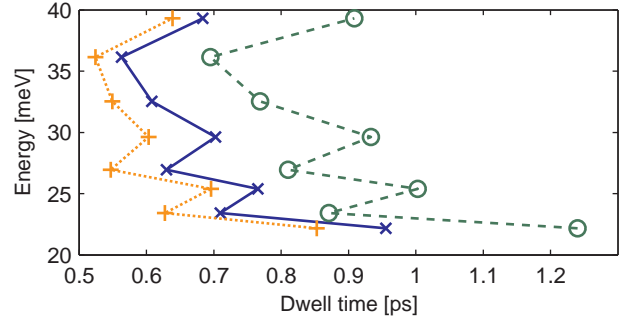}
\caption{(Color online) Simulation results for electron dwell times, as
obtained by fully taking into account (x-marks, solid curves) and ignoring the
exchange effect (plus signs, dotted curves), and by ignoring parallel spin
collisions (circles, dashed curves). The lines are guide to the eye.}%
\label{dwell_spin}%
\end{figure}

In Fig. \ref{pT_spin}, the relative occupations and the electron temperatures
are shown at $T_{\mathrm{L}}=10\,\mathrm{K}$ for the eight energy levels
within a miniband, characterized by their corresponding eigenenergies (compare
Fig. \ref{cond}(a)). The electron temperatures are extracted from the
(generally non-thermal) carrier distributions in each subband by a least
square fit. While the subband occupations do not depend much on the
implementation of the exchange effect, the electron temperatures show a
stronger dependence. For example, the extracted temperature for the $8$th
level (at $39.3\,\mathrm{meV}$) varies from $121.6\,\mathrm{K}$ to
$146.7\,\mathrm{K}$, depending on the implementation of the spin dependence.
The relative insensitivity of the population and the strong dependence of the
electron temperature and gain on the implementation can be understood by
looking at the average dwell time of an electron in a given subband, which is
the inverse of the outscattering rate from this subband. In Fig.
\ref{dwell_spin}, the dwell time is shown for the eight energy levels, again
characterized by their eigenenergies. The scattering is lowest, i.e., the
dwell time is highest, when parallel spin collisions are ignored in the
simulation (dashed curve). This is consistent with the reduced bandwidth and
enhanced peak value of the corresponding gain profile in Fig. \ref{g_spin}(a),
which are directly related to the outscattering
rate.\cite{2009JAP...105l3102J} On the other hand, the electron dwell time
increases by a similar factor for all the subbands, thus explaining the
relative insensitivity of subband occupations to the chosen implementation of
Eq. (\ref{M}). 

\subsection{Screening}

Due to the computational effort involved in the RPA, screening is commonly
taken into account using simplified models rather than solving Eq.
(\ref{Vscr}) directly. The screening can for instance be considered by
introducing a screening wavenumber $q_{\mathrm{s}}$ in Eq. (\ref{Vbare}),
i.e., replacing the prefactor $e^{2}/\left(  2\varepsilon Q\right)  $ by
$e^{2}/\left[  2\varepsilon\left(  Q+q_{\mathrm{s}}\right)  \right]  $. In
single subband models, $q_{\mathrm{s}}$ is obtained from Eq. (\ref{Vscr}) by
assuming that screening is caused only by a single subband, e.g., the ground
state.\cite{2005JAP....97d3702B,2006ApPhL..89u1115L} A modified approach,
which has been shown to yield improved results for RP structures, is the
modified single subband model\cite{2006ApPhL..89u1115L} with%
\begin{equation}
q_{\mathrm{s}}=\frac{e^{2}}{2\varepsilon}\frac{m^{\ast}}{\pi\hbar^{2}}\sum
_{i}f_{i}\left(  \mathbf{k}=\mathbf{0}\right)  , \label{qs}%
\end{equation}
where $i$ sums over the subbands in one period.\begin{figure}[ptb]
\includegraphics{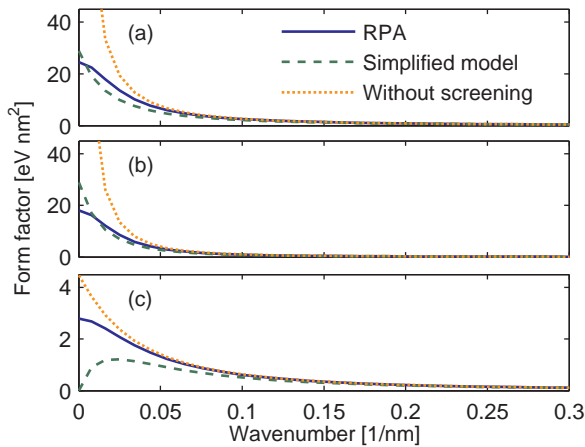}
\caption{(Color online) Screened and unscreened Coulomb matrix elements for
the structure shown in Fig. \ref{cond}(a). (a) $V_{1111}$; (b) $V_{1212}$; (c)
$V_{1122}$.}%
\label{V}%
\end{figure}

In Fig. \ref{V}, the intrasubband Coulomb matrix elements $V_{1111}$ (Fig.
\ref{V}(a)) and $V_{1212}$ (Fig. \ref{V}(b)), as well as the intersubband
element $V_{1122}$\ (Fig. \ref{V}(c)) are shown as a function of the
wavenumber $Q$. Here, $1$ and $2$ denote the upper laser level at
$22.2\,\mathrm{meV}$ and the level directly above at $23.4\,\mathrm{meV}$, see
Fig. \ref{cond}(a). Displayed are the screened matrix elements based on the
RPA (solid lines) and the simplified model according to Eq. (\ref{qs}) with
$q_{\mathrm{s}}=0.0237\,\mathrm{nm}^{-1}$ (dashed lines), as well as the bare
matrix elements defined in Eq. (\ref{Vbare}) (dotted lines). As can be seen
from Fig. \ref{V}(c), in the simplified screening model the intersubband
elements approach zero for small wavenumbers, in contrast to the exact
implementation of the RPA. A better approach consists in applying the
simplified screening model only to the intrasubband matrix elements, and to
treat intersubband scattering as unscreened.\cite{2007JAP...101f3101G}%
\begin{figure}[ptb]
\includegraphics{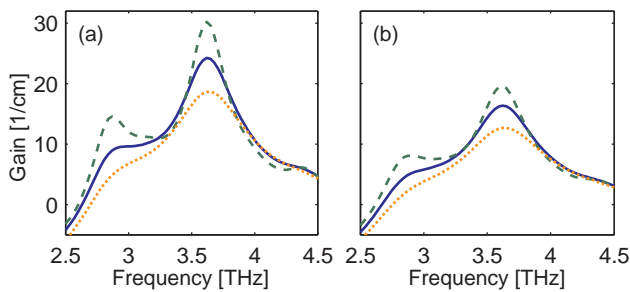}
\caption{(Color online) Simulated spectral gain vs frequency, as obtained by
taking into account screening in the RPA (solid curves), and using the
modified single subband model for all matrix elements (dashed curves) or for
the intrasubband elements only (dotted curves), i.e., treating intersubband
elements as unscreened. (a) $T_{\mathrm{L}}=10\,\mathrm{K}$; (b)
$T_{\mathrm{L}}=90\,\mathrm{K}$.}%
\label{g_screen}%
\end{figure}

\begin{figure}[ptbptb]
\includegraphics{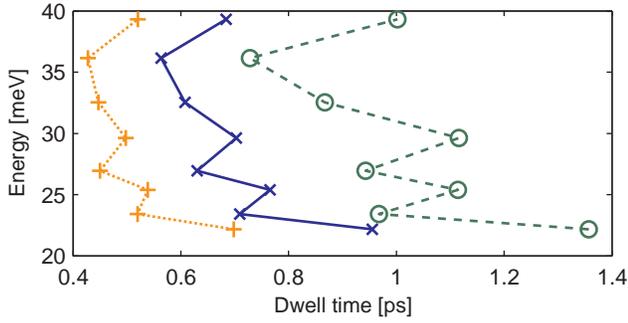}
\caption{(Color online) Simulation results for electron dwell times, as
obtained by taking into account screening in the RPA (x-marks, solid curves),
and using the modified single subband model for all matrix elements (circles,
dashed curves) or for the intrasubband elements only (plus signs, dotted
curves). The lines are guide to the eye.}%
\label{dwell_screen}%
\end{figure}

Figure \ref{g_screen} contains the self-self-consistently simulated gain
spectra at $T_{\mathrm{L}}=10\,\mathrm{K}$ and $T_{\mathrm{L}}=90\,\mathrm{K}%
$. In contrast to Fig. \ref{g_spin}, the exchange effect is included, but
different screening models are used. The reference curve based on the exact
evaluation of the RPA (solid curves) agrees well with experiment, see Section
\ref{space_charge}. Applying the simplified screening model to all matrix
elements overestimates the screening of the intersubband elements, compare
Fig. \ref{V}(c), and thus results in an underestimation of scattering. The
resulting spectral gain profile at $10\,\mathrm{K}$ (dashed curve in Fig.
\ref{g_screen}(a)) features a $25\%$ enhanced gain peak and an excessively
narrow FWHM\ width of $0.43\,\mathrm{THz}$, as compared to an experimental
value of $0.85\,\mathrm{THz}$. On the other hand, completely ignoring the
screening effect for the intersubband matrix elements overestimates the
intersubband scattering, thus resulting in a lowered and \ broadened gain
profile (dotted curves). As can be seen from Fig. \ref{g_screen}(b), the
simulation results at $T_{\mathrm{L}}=90\,\mathrm{K}$ are affected in a
similar way by using above discussed approximations. \begin{figure}[ptb]
\includegraphics{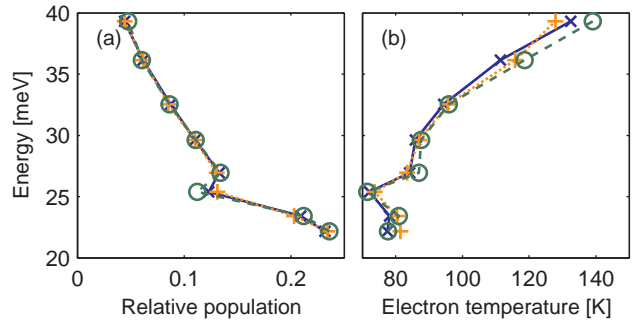}
\caption{(Color online) Simulation results for (a) relative subband
occupations and (b) subband temperatures, as obtained by taking into account
screening in the RPA (x-marks, solid curves), and using the modified single
subband model for all matrix elements (circles, dashed curves) or for the
intrasubband elements only (plus signs, dotted curves). The lines are guide to
the eye.}%
\label{pT_screen}%
\end{figure}In Fig. \ref{pT_screen}, the obtained relative subband occupations
and electron temperatures are compared at $T_{\mathrm{L}}=10\,\mathrm{K}$ for
the different implementations of screening. Although the simulated gain in
Fig. \ref{g_screen}\ greatly depends on the applied screening model, the
occupations of the miniband levels in Fig. \ref{pT_screen}(a) are quite
insensitive to the chosen implementation, similarly as in Fig. \ref{pT_spin}.
This is again consistent with the fact that the average dwell time of an
electron in a level, shown in Fig. \ref{dwell_screen}, strongly depends on the
chosen screening model, but is changed by a similar factor for all subbands.
We attribute this to the quasi-continuous nature of the minibands in the BTC
structure. Contrariwise, for RP QCLs, where the energetic separation of the
levels greatly varies, simple approaches like the single subband screening
model have been shown to considerably affect the simulation results for the
level occupations.\cite{2005JAP....97d3702B,2006ApPhL..89u1115L} Here, only a
single subband, typically the ground level, is
considered.\cite{2006ApPhL..89u1115L} However, it should be mentioned that
also for the investigated BTC structure, this approach would yield somewhat
less accurate results than evaluating Eq. (\ref{qs}), as done in our
simulations. The kinetic electron distribution within each subband,
represented by fitted electron temperatures in Fig. \ref{pT_screen}(b), shows
a moderate dependence on the screening model. For example, the extracted
temperatures in the $8$th level (at $39.3\,\mathrm{meV}$) ranges between
$127.9\,\mathrm{K}$ and $139.0\,\mathrm{K}$ for the different screening
models. Here again, more significant deviations from the RPA result would be
obtained by using a single subband screening model rather than Eq. (\ref{qs}).

\section{CONCLUSIONS}

Employing an EMC analysis, we have investigated the role of Coulomb
interactions in THz BTC structures, and assessed the validity of common
approximations for modeling Coulomb effects. We have shown that space charge
effects lead to considerable conduction band bending, having a significant
influence on the obtained wave functions and the energy level alignment. These
mechanisms are properly accounted for in a self-self-consistent approach,
i.e., by iteratively performing SP and EMC simulations until mutual
convergence is achieved. Ignoring these effects leads to excessively lowered
and broadened gain profiles, in sharp contradiction to experimental results.
Also e-e scattering between the closely spaced miniband levels plays an
important role for the carrier transport and the spectral gain profile. Common
approximations in implementing e-e scattering, like neglecting the
spin-related exchange effect or using simplified screening models, results in
considerable deviations for the obtained gain profile. Even refined
approximations, like the modified single subband screening model, introduce an
error of approximately $25\%$ for the peak gain in the investigated structure,
as compared to simulations based on the full RPA. Combined with ignoring the
spin dependence, the deviation can easily exceed $30\%$. With space charge
effects and e-e scattering properly accounted for, the EMC analysis is shown
to yield meaningful results for the investigated BTC design, which are found
to be in good agreement with experiment.%

\begin{acknowledgments}
C.J. and A.M. acknowledge support from the Emmy Noether program of the German
Research Foundation Grant No. DFG, JI115/1-1.
\end{acknowledgments}


\newpage

\begin{thebibliography}{9}                                                                                                %

\bibitem {2005ApPhL..86u1117J}V.~D. {Jovanovi{\'{c}}}, D.~{Indjin},
N.~{Vukmirovi{\'{c}}}, Z.~{Ikoni{\'{c}}}, P.~{Harrison}, E.~H. {Linfield},
H.~{Page}, X.~{Marcadet}, C.~{Sirtori}, C.~{Worrall}, H.~E. {Beere}, and D.~A.
{Ritchie}, \newblock Appl. Phys. Lett. \textbf{86}, 211117 (2005).

\bibitem {2009PhRvB..79s5323K}T.~{Kubis}, C.~{Yeh}, P.~{Vogl}, A.~{Benz},
G.~{Fasching}, and C.~{Deutsch}, \newblock Phys. Rev. B \textbf{79}, 195323 (2009).

\bibitem {2001ApPhL..78.2902I}R.~C. {Iotti} and F.~{Rossi}, \newblock Appl.
Phys. Lett. \textbf{78}, 2902 (2001).

\bibitem {2003ApPhL..83..207C}H.~{Callebaut}, S.~{Kumar}, B.~S. {Williams},
Q.~{Hu}, and J.~L. {Reno}, \newblock Appl. Phys. Lett. \textbf{83}, 207 (2003).

\bibitem {2004ApPhL..84..645C}H.~{Callebaut}, S.~{Kumar}, B.~S. {Williams},
Q.~{Hu}, and J.~L. {Reno}, \newblock Appl. Phys. Lett. \textbf{84}, 645 (2004).

\bibitem {2005JAP....97d3702B}O.~{Bonno}, J.-L. {Thobel}, and F.~{Dessenne},
\newblock J. Appl. Phys. \textbf{97}, 043702 (2005).

\bibitem {2006ApPhL..89u1115L}J.~T. {L{\"{u}}} and J.~C. {Cao}, \newblock
Appl. Phys. Lett. \textbf{89}, 211115 (2006).

\bibitem {2007JAP...101h6109J}C.~{Jirauschek}, G.~{Scarpa}, P.~{Lugli}, M.~S.
{Vitiello}, and G.~{Scamarcio}, \newblock J. Appl. Phys. \textbf{101}, 086109 (2007).

\bibitem {2008pssc..5..221J}C.~{Jirauschek} and P.~{Lugli}, \newblock Phys.
Status Solidi C \textbf{5}, 221 (2008).

\bibitem {2001ApPhL..79.3920K}R.~{K{\"{o}}hler}, R.~C. {Iotti},
A.~{Tredicucci}, and F.~{Rossi}, \newblock Appl. Phys. Lett. \textbf{79}, 3920 (2001).

\bibitem {2009JAP...105l3102J}C.~{Jirauschek} and P.~{Lugli}, \newblock J.
Appl. Phys. \textbf{105}, 123102 (2009).

\bibitem {2006ApPhL..88f1119L}J.~T. {L{\"{u}}} and J.~C. {Cao}, \newblock
Appl. Phys. Lett. \textbf{88}, 061119 (2006).

\bibitem {1985Lugli_excl}P.~{Lugli} and D.~K. {Ferry}, \newblock IEEE Trans.
Electron Devices \textbf{32}, 2431 (1985).

\bibitem {2009IJQE...45..1059J}C.~{Jirauschek}, \newblock IEEE J. Quantum
Electron. \textbf{45}, 1059 (2009).

\bibitem {1988PhRvB..37.2578G}S.~M. {Goodnick} and P.~{Lugli}, \newblock
Phys. Rev. B \textbf{37}, 2578 (1988).

\bibitem {1995PhRvB..5116860M}M.~{Mo{\v s}ko}, A.~{Mo{\v s}kov{\'a}}, and
V.~{Cambel}, \newblock Phys. Rev. B \textbf{51}, 16860 (1995).

\bibitem {1999PhRvB..5915796L}S.-C. {Lee} and I.~{Galbraith}, \newblock  Phys.
Rev. B \textbf{59}, 15796 (1999).

\bibitem {1994SeScT...9..478M}M.~{Mo{\v{s}}ko} and A.~{Mo{\v{s}}kov{\'{a}}},
\newblock Semicond. Sci. Technol. \textbf{9}, 478 (1994).

\bibitem {2003ApPhL..82.3165S}G.~{Scalari}, L.~{Ajili}, J.~{Faist},
H.~{Beere}, E.~{Linfield}, D.~{Ritchie}, and G.~{Davies}, \newblock Appl.
Phys. Lett. \textbf{82}, 3165 (2003).

\bibitem {2006PhRvB..73h5311L}A.~{Leuliet}, A.~{Vasanelli}, A.~{Wade},
G.~{Fedorov}, D.~{Smirnov}, G.~{Bastard}, and C.~{Sirtori}, \newblock Phys.
Rev. B \textbf{73}, 085311 (2006).

\bibitem {2008ApPhL..92h1102N}R.~{Nelander} and A.~{Wacker}, \newblock Appl.
Phys. Lett. \textbf{92}, 081102 (2008).

\bibitem {2007JAP...101f3101G}X.~{Gao}, D.~{Botez}, and I.~{Knezevic},
\newblock J. Appl. Phys. \textbf{101}, 063101 (2007).
\end{thebibliography}
\end{document}